\documentclass[aps,pra,superscriptaddress,twocolumn]{revtex4}
\usepackage{amsmath}
\usepackage{amssymb}
\usepackage{graphicx}
\usepackage{dcolumn}

\begin{document}

\title{The influence of device geometry\\
on many-body effects in quantum point contacts:\\
Signatures of the 0.7 anomaly, exchange and Kondo}


\author{E.~J.~Koop}
\email[e-mail: ]{e.j.koop@rug.nl}
\author{A.~I.~Lerescu}
\author{J.~Liu}
\author{B.~J.~van~Wees}
\affiliation{Physics of Nanodevices Group, Zernike Institute for Advanced Materials,\\
University of Groningen, Nijenborgh 4, 9747 AG  Groningen, The
Netherlands}
\author{D.~Reuter}
\author{A.~D.~Wieck}
\affiliation{Angewandte Festk\"{o}rperphysik, Ruhr-Universit\"{a}t
Bochum, D-44780 Bochum, Germany}
\author{C.~H.~van~der~Wal}
\affiliation{Physics of Nanodevices Group, Zernike Institute for Advanced Materials,\\
University of Groningen, Nijenborgh 4, 9747 AG  Groningen, The
Netherlands}

\date{\today}

\vskip 10mm

\begin{abstract}

The conductance of a quantum point contact (QPC) shows several
features that result from many-body electron interactions. The spin
degeneracy in zero magnetic field appears to be spontaneously lifted
due to the so-called 0.7 anomaly. Further, the g-factor for
electrons in the QPC is enhanced, and a zero-bias peak in the
conductance points to similarities with transport through a Kondo
impurity. We report here how these many-body effects depend on QPC
geometry. We find a clear relation between the enhanced g-factor and
the subband spacing in our QPCs, and can relate this to the device
geometry with electrostatic modeling of the QPC potential. We also
measured the zero-field energy splitting related to the 0.7 anomaly,
and studied how it evolves into a splitting that is the sum of the
Zeeman effect and a field-independent exchange contribution when
applying a magnetic field. While this exchange contribution shows
sample-to-sample fluctuations and no clear dependence on QPC
geometry, it is for all QPCs correlated with the zero-field
splitting of the 0.7 anomaly. This provides evidence that the
splitting of the 0.7 anomaly is dominated by this field-independent
exchange splitting. Signatures of the Kondo effect also show no
regular dependence on QPC geometry, but are possibly correlated with
splitting of the 0.7 anomaly.

\end{abstract}

\maketitle

\section{Introduction}

A quantum point contact (QPC) is a short channel that carries
ballistic one-dimensional electron transport between two reservoirs.
Its conductance as a function of channel width is quantized
\cite{vanWees1988,Wharam1988} and shows plateaus at integer
multiples of $2e^2/h$, where $e$ is the electron charge and $h$
Planck's constant. This quantization of the conductance can be
understood with a noninteracting electron picture. However, there
are several features in the conductance that result from many-body
interaction effects between electrons. The effective electron
g-factor is enhanced and almost all semiconductor QPCs show an
additional plateau at $\sim 0.7(2e^2/h)$, the so-called 0.7 anomaly.
Further, electron transport through QPCs tuned to conditions where
the 0.7 anomaly appears has similarities with transport through a
Kondo impurity. These many-body effects are not yet fully
understood, and in particular understanding the 0.7 anomaly has been
the topic of on-going research for more than a decade now
 \cite{Fitzgerald2002,Ennslin2006}. A consistent picture of these
effects is of interest for spintronics and quantum information
proposals where QPCs are a key element, and QPCs are now also a key
model system for studies of many-body physics in nanodevices.

\begin{figure}
  \includegraphics[width=0.8\columnwidth]{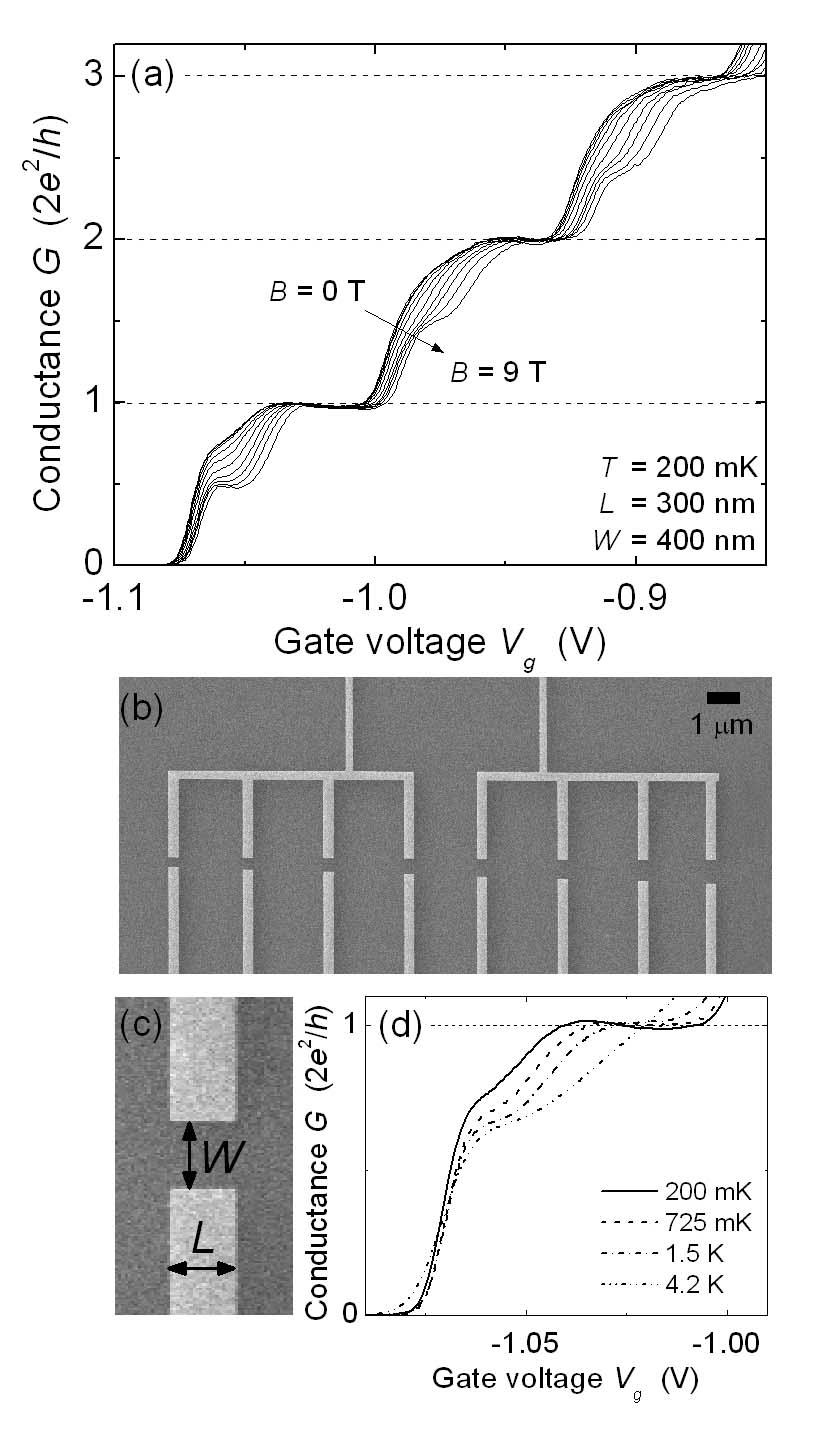}\\
  \caption{(a) The differential
conductance $G$ as a function of gate voltage $V_g$ at 200 mK,
  for a QPC with $L$ = 300 nm and $W$ = 400 nm. The in-plane magnetic field is
  increased from $B = 0 \; {\rm T}$ to $B = 9 \; {\rm T}$. The first three spin-degenerate
  plateaus at integer multiples of $2e^2/h$ for $B = 0 \; {\rm T}$ split into six spin-resolved plateaus
  integer multiples of $e^2/h$ for $B = 9 \; {\rm T}$. (b)
  Micrograph of a device containing 8 QPCs. From left to right the
  width $W$ is increased, where $W$ is defined as the spacing between the gate electrodes as shown in (c). $L$ is the
  length of the channel. Table \ref{TABLE:SampleWandL} contains all values for $L$ and $W$ of the measured devices. (d) Differential conductance $G$ as a function of gate voltage $V_g$ at zero field
  for different temperatures. The 0.7 anomaly becomes more pronounced with increasing temperature.}
  \label{FIG:DeviceAndConductance}
\end{figure}


Several models have been proposed that relate the 0.7 anomaly to a
spontaneous spin splitting in zero magnetic field \cite{Thomas1996,
Thomas1998, Kristensen2000, Meir2002, Starikov2003, Reilly2005,
DiCarlo2006}, since the 0.7 plateau evolves continuously into the
spin-resolved plateau at 0.5 (2e$^2$/h) when an in-plane magnetic
field is applied. A recent theory paper \cite{Rejec2006} presented
spin-density functional calculations of realistic QPC geometries
that show that a localized state can exist near pinch-off in a QPC,
providing a theoretical background for the Kondo-like physics that
was found experimentally \cite{Cronenwett2002}. Other studies have
proposed electron-phonon scattering \cite{Matveev2003}, Wigner
crystal formation \cite{Matveev2004}, or a dynamical Coulomb
blockade effect \cite{Bulka2007} as the microscopic origin of the
0.7 anomaly. Graham \textit{et al.} reported evidence that many-body
effects also play a role in magnetic fields at crossings between
Zeeman levels of different subbands \cite{Graham2003}, and at
crossings of spin-split subbands with reservoir levels
 \cite{Graham2005}.

We report here how these many-body effects in QPCs depend on the QPC
geometry. We study the energy spacing between the one-dimensional
subbands and spin-splittings within one-dimensional subbands, both
in zero field and high magnetic fields. While this type of data from
individual devices has been reported before
 \cite{Thomas1998,Cronenwett2001}, we report here data from a set of
12 QPCs with identical material parameters. Our measurements show a
clear correlation between the subband spacing $\hbar \omega_{12}$
and the enhancement of the effective g-factor $|g^*|$. Both also
depend in a regular manner on the geometry of the QPC, and we can
understand this behavior using electrostatic modeling of the QPC
potential.

The appearance of the 0.7 anomaly and signatures of the Kondo effect
do not show a regular dependence on QPC geometry. Intriguingly,
however, we find that in high magnetic fields there is a
field-independent exchange contribution to the spin-splitting for
the lowest one-dimensional subband in addition to the regular Zeeman
splitting, and this exchange contribution is clearly correlated with
the zero-field splitting of the 0.7 anomaly. However we do not claim
that this zero-field splitting leads to a static ferromagnetic
polarization. This new observation provides evidence that the
splitting of the 0.7 anomaly is dominated by this field-independent
exchange splitting. The Kondo effect appears as a zero-bias peak in
the differential conductance $G$, and the width of this peak is set
by the Kondo temperature $T_{K}$, an energy scale that represents
the strength of the Kondo effect. Our measurements of $T_{K}$
suggest a correlation between $T_{K}$ and the splitting of the 0.7
anomaly.

This paper is organized as follows. Section \ref{Sect:ExpDetails}
presents information about sample fabrication and measurement
techniques. In section \ref{Sect:EnergySplittings} we present
measurements of the conductance of our set of QPCs, and we extract
the energy splittings between subbands and spin splittings. In
section \ref{Sect:ManyBodyEffects} we focus on analyzing the
signatures of many-body effects in our QPC data, before ending with
concluding remarks in the last section.


\section{Experimental realization}\label{Sect:ExpDetails}

Our devices were fabricated using a ${\rm GaAs}/{\rm Al}_{0.32}{\rm
Ga}_{0.68}{\rm As}$ heterostructure with a 2DEG at 114 nm below the
surface from modulation doping with Si. The buffer layer had a
thickness of 36.8 nm, and Si doping was about $n_{Si} \approx 1
\cdot 10^{24} \; {\rm m^{-3} }$.  At 4.2 K, the mobility of the 2DEG
was $\mu = 159 \; {\rm m^{2}/Vs }$, and the electron density $n_{s}
= (1.5 \pm 0.1) \cdot 10^{15} \; {\rm m^{-2} }$. A QPC is formed by
applying a negative gate voltage $V_{g}$ to a pair of electrodes on
the wafer surface. The 2DEG below the electrodes is then fully
depleted, and tuning of $V_g$ allows for controlling the width of a
short one-dimensional transport channel. Our QPCs had different
values for the length $L$ and width $W$ for the electrode spacing
that defines the device (see Table \ref{TABLE:SampleWandL}, and
Figs.~\ref{FIG:DeviceAndConductance}b,c). Note that $W$ should not
be confused with the actual width of the transport channel that is
controlled with $V_{g}$. The depletion gates were defined with
standard electron-beam lithography and lift-off techniques, using
deposition of 15 nm of Au with a Ti sticking layer. The reservoirs
were connected to macroscopic leads via Ohmic contacts, which were
realized by annealing a thin Au/Ge/Ni layer that was deposited on
the surface.

All QPCs were fabricated in close proximity of each other on a
single central part of the wafer to ensure the same heterostructure
properties for all QPCs. The set of 8 QPCs for which we varied $L$
(Device 1 in Table \ref{TABLE:SampleWandL}) had all QPCs within a
range of about $10 \; {\rm \mu m}$. The set of 8 QPCs for which we
varied $W$ (Device 2 in Table \ref{TABLE:SampleWandL} and
Fig.~\ref{FIG:DeviceAndConductance}b) had an identical layout, and
was positioned at 2 mm from Device 1. Thus, all semiconductor
processing steps (resist spinning, e-beam lithography, metal
deposition, etc.) could be kept nominally identical for all 16 QPCs.
Electron-microscope inspection of the measured devices (after the
measurements) confirmed that the dimensions of all gate electrodes
were within 10 nm of the designed values (see table
\ref{TABLE:SampleWandL}. In our data this appears as a very regular
dependence of QPC properties (see for example the discussion of the
pinch-off voltage $V_{po}$ and subband spacing $\hbar \omega_{12}$
in the next section) on $L$ and $W$ for QPCs within the sets of
Device 1 and 2. At the same time, two devices from two different
sets with nominally identical values of $L$ and $W$ (labeled (1) and
(2) in Figs.~\ref{FIG:PinchOffAndSubbandSpacing} and
\ref{FIG:GfactorAndEnergySplittingVSShape}) show slightly different
QPC properties (in particular for the subband spacing $\hbar
\omega_{12}$). This is not fully understood.

\begin{table}
\vspace{3mm}
\begin{tabular}{|c|c|c|c|c|c|c|c|c|}
\multicolumn{9}{l}{Device 1}\\
\hline
$L$ (nm) & 100 & 150 & 200 & 250 & 300 & 350 & 400 & 450 \\
$W$ (nm) & 350 & 350 & 350 & 350 & 350 & 350 & 350 & 350 \\
\hline
\multicolumn{9}{l}{Device 2}\\
\hline
$L$ (nm) & 300 & 300 & 300 & 300 & 300 & 300 & 300 & 300 \\
$W$ (nm) & 200 & 250 & 300 & 350 & 400 & 450 & 500 & 550 \\
\hline
\end{tabular}
\caption {Dimensions of the measured QPCs. The QPC length $L$ and
width $W$ are defined as in Fig.~\ref{FIG:DeviceAndConductance}c.}
\label{TABLE:SampleWandL} \vspace{3mm}
\end{table}

Measurements were performed in a dilution refrigerator with the
sample at temperatures from $\sim 5 \; {\rm mK}$ to 4.2~K. For all
our data the temperature dependence saturated when cooling below
$\sim 200$~mK. We therefore assume for this report that this is the
lowest effective electron temperature that could be achieved. For
measuring the differential conductance $G$ we used standard lock-in
techniques at 380 Hz, with an ac voltage bias $V_{ac}=10 \; {\rm \mu
V}$. Only the $V_{-}$ contact was connected to the grounded
shielding of our setup, and all gate voltages were applied with
respect to this ground. The in-plane magnetic field was applied
perpendicular to the current direction, and the current in the QPCs
was along the $[110]$ crystal orientation. Alignment of the sample
with the magnetic field was within 1$^\circ$, as determined from
Hall voltage measurements on the 2DEG. We have data from 12
different QPCs from the set of 16 that we cooled down. From these
QPCs 4 could not be measured. For two this was due to the presence
of strong telegraph noise in conductance signals. Two other QPCs did
not show clear conductance plateaus.

For analyzing QPC conductance values we subtracted a magnetic field
and temperature dependent series resistance (from the wiring and
filters, Ohmic contacts and 2DEG) from the transport data that was
obtained with a voltage-bias approach. The criterium here was to
make the observed conductance plateaus coincide with integer
multiples of $2e^{2}/h$ or $e^{2}/h$.


\section{Spin splitting and energy splitting between QPC subbands} \label{Sect:EnergySplittings}

\subsection{QPC conductance and energy splittings}

Figure~\ref{FIG:DeviceAndConductance}a presents the differential
conductance $G$ of a QPC as a function of $V_{g}$, with the
source-drain voltage $V_{sd} \approx 0$. Increasing $V_{g}$ from
pinch-off ($G=0$) lowers and widens the saddle-point-like potential
that defines the short transport channel. Consequently, an
increasing number of one-dimensional subbands gets energies below
the Fermi level. In zero magnetic field, this results in a step of
$2e^2/h$ in the conductance each time an additional subband starts
to contribute to transport. We label these spin-degenerate subbands
with a number $N$, starting with $N=1$ for the lowest subband. With
a high in-plane magnetic field $B$ the spin degeneracy within each
subband $N=1,2,3...$ is lifted, and the conductance increases now in
steps of $e^2/h$.

\begin{figure}[h!]
  \includegraphics[width=0.95\columnwidth]{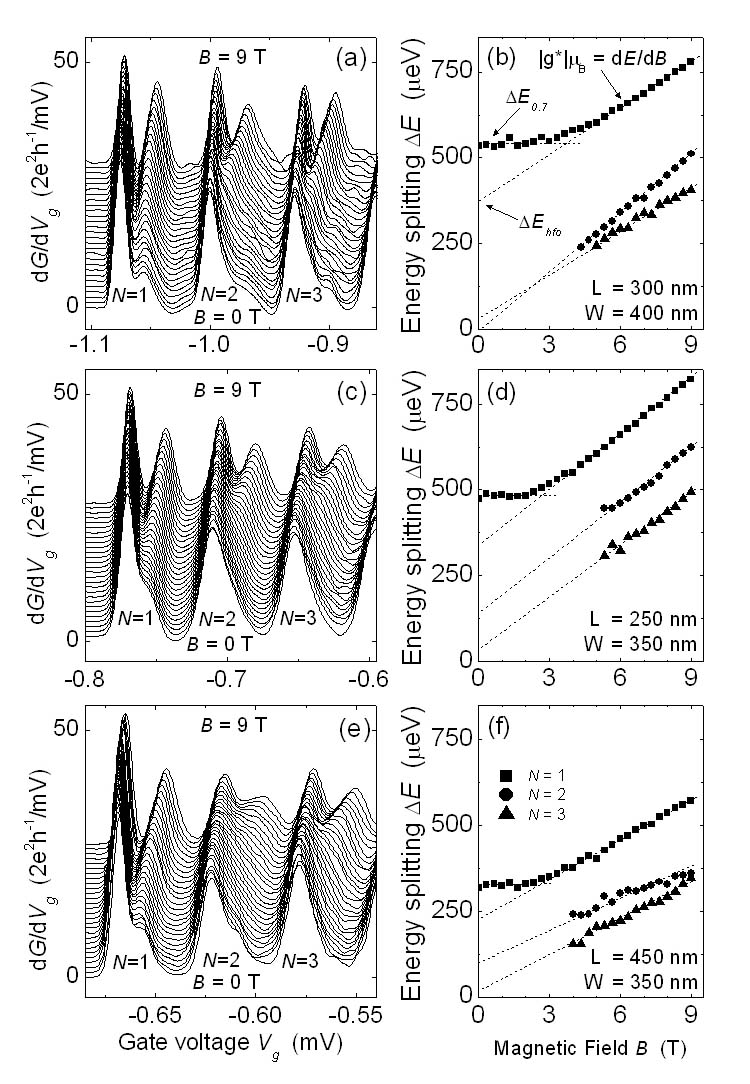}\\
  \caption{(a) Transconductance ${\rm d}G / {\rm d}V_g$ traces (offset
vertically) obtained from the data in Fig.
\ref{FIG:DeviceAndConductance}a (from the QPC with $L$ = 300 nm and
$W$ = 400 nm). The 0.7 anomaly appears as a splitting of the
transconductance peak for the $N=1$ subband at $B = 0 \; {\rm
  T}$.  (b) Energy splittings $\Delta E$ obtained from
  the transconductance traces in (a),
  as a function of magnetic field. The traces present $\Delta E$ for the subbands $N=1,2,3$, see the legend in (f).
  These $\Delta E$ traces are characterized (results presented in Fig.~\ref{FIG:GfactorAndEnergySplittingVSShape})
  with two or three parameters for each subband $N=1,2,3$: An effective g-factor
  $|g^*|$, the offset from a linear Zeeman effect in high fields, characterized by the
  high-field offset $\Delta E_{hfo}$, and for $N=1$ at low fields
  the   energy splitting of the 0.7 anomaly, $\Delta E_{0.7}$. See
  text for details. (c)-(f) Transconductance traces ${\rm d}G / {\rm d}V_g$ and energy splittings $\Delta E$ as in (a), (b)
  obtained for a QPC with $L$~=~250 nm and $W$~=~350 nm in (c), (d) and $L$~=~450 nm and $W$~=~350 nm in (e),
  (f). All data from measurements at 200~mK}
  \label{FIG:TransconductanceAndEnergySplitting}
\end{figure}


We use this type of data to determine the energy splitting $\Delta
E$ between spin-up and spin-down levels within the subbands
$N=1,2,3$, and the spacing $\hbar \omega_{12}$ between the $N=1$ and
$N=2$ subband (a measure for the degree of transverse confinement in
the channel). The onset of transport through a next (spin-polarized)
subband appears as a peak in transconductance (${\rm d}G / {\rm
d}V_g$) traces as in
Figs.~\ref{FIG:TransconductanceAndEnergySplitting}a,c,e, which we
derive from traces as in Fig.~\ref{FIG:DeviceAndConductance}a. We
assume that each subband contributes in a parallel manner to the QPC
conductance, and the transconductance curves can then be analyzed as
a superposition of peaks, with one (two) peak(s) per (spin-split)
subband. We then determine the peak spacings $\Delta V_{g}$ along
the $V_{g}$ axis by fitting one or two peaks per subband on the
transconductance traces (using least squares fitting with a Gaussian
peak shape). The specific shape of a step between the quantized
conductance plateaus depends on the shape of the saddle-point-like
potential that defines the QPC \cite{Buttiker1990}, and can result
in asymmetric transconductance peaks. We checked that this is not a
significant effect for our analysis.

Subsequently, transconductance data (not shown) from nonlinear
transport measurements is used for converting $\Delta V_{g}$ values
into energy splittings \cite{Patel1991}. Here, the onsets of
conductance plateaus appear as diamond shaped patterns in the
$V_{sd} - V_{g}$ plane. The width of these diamonds along the
$V_{sd}$ axis defines the subband spacing, and we use this to
determine the spacing $\hbar \omega_{12}$ between the $N=1$ and
$N=2$ subband. The slopes of the diamonds can be used to convert a
gate-voltage scale into energy scale \cite{Patel1991}. In this
analysis of $\hbar \omega_{12}$ and conversion of $\Delta V_{g}$
into spin splittings $\Delta E$ we observed a weak dependence on
magnetic field and temperature, and took this in account.

The 0.7 anomaly is clearly visible in the data set presented in
Fig.~\ref{FIG:DeviceAndConductance}. The conductance trace for zero
field in Fig.~\ref{FIG:DeviceAndConductance}a shows besides
pronounced steps of $2e^2/h$ an additional shoulder at $\sim
0.7(2e^2/h)$, which becomes more pronounced at higher temperatures
(Fig.~\ref{FIG:DeviceAndConductance}d). With increasing magnetic
field, the 0.7 anomaly evolves into the first spin-resolved
conductance plateau at $e^2/h$. In
Figs.~\ref{FIG:TransconductanceAndEnergySplitting}a,c,e the 0.7
anomaly appears as a zero-field splitting in the transconductance
peak for $N=1$, which evolves into two spin-split peaks in high
fields. In earlier work this observation was the basis for assuming
that the 0.7 anomaly results from a spontaneous removal of spin
degeneracy in zero field \cite{Thomas1996,Thomas1998}. For our
analysis here we assume that the 0.7 anomaly is indeed related to
such a spontaneous spin splitting for the first subband.  In high
fields, all 12 QPCs showed also for $N=2$ and higher a pronounced
spin splitting into two transconductance peaks, but these subbands
did not clearly show a zero-field splitting. We emphasize again that
we do not claim there is a static ferromagnetic polarization due to
this splitting.

\begin{figure}
  \includegraphics[width=1\columnwidth]{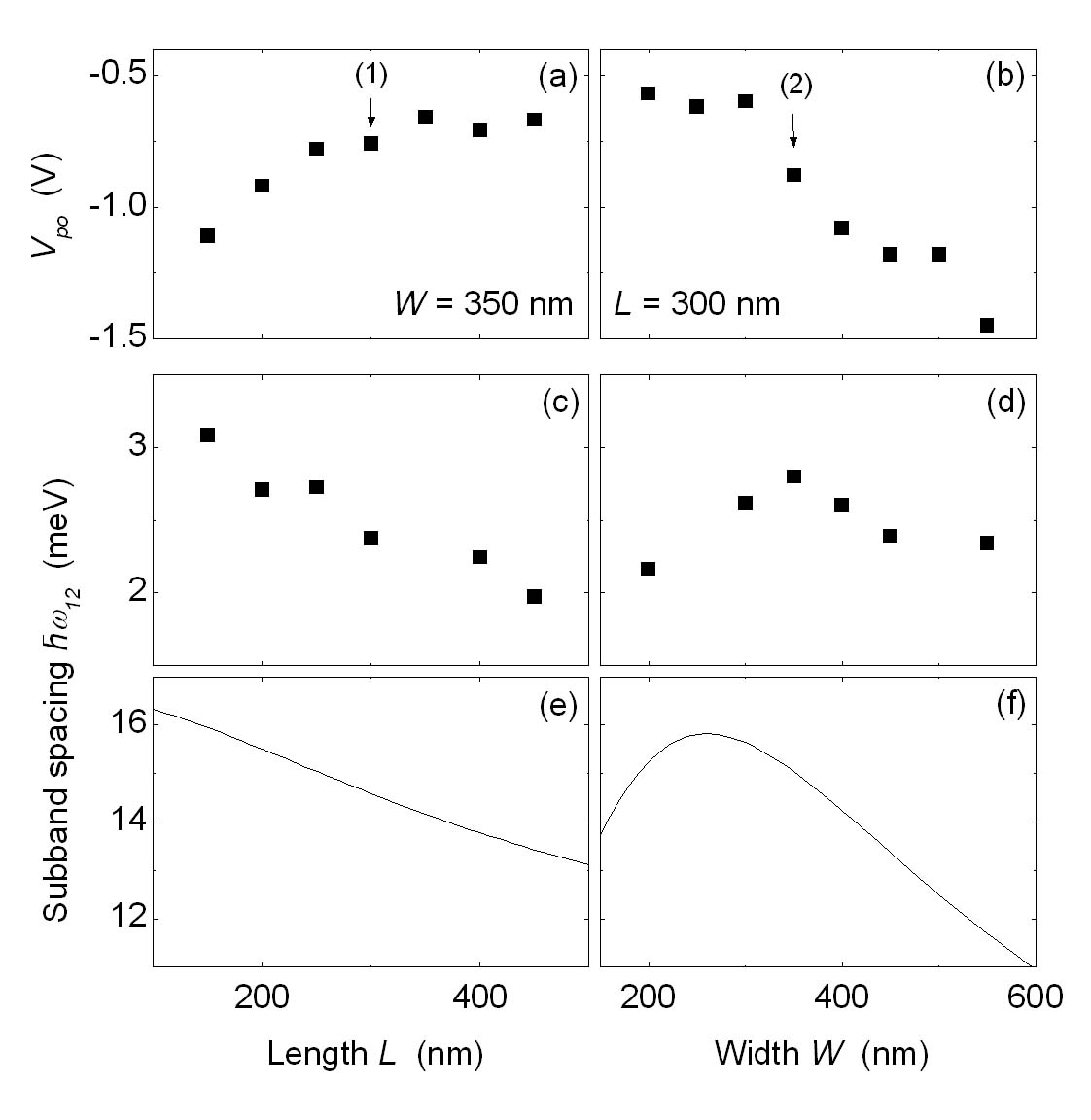}\\
  \caption{(a),(b) The
pinch-off voltage $V_{po}$ as a function of QPC
  length $L$ (with fixed width $W$ = 350 nm), and as a function QPC width $W$ (with
  fixed length $L$ = 300 nm).
Data points labeled with (1) and (2) are from two different devices
with nominally identical values for $L$ and $W$ (see text for
details).
  (c),(d) The measured subband spacing $\hbar
  \omega_{12}$ as a function of $L$ and $W$.
  (e),(f) Calculated subband spacing $\hbar
  \omega_{12}$ as a function of $L$ and $W$, from electrostatic modeling of the QPC potential. The results qualitatively reproduce
  the trend in the experimental data in (c),(d).}
 \label{FIG:PinchOffAndSubbandSpacing}
\end{figure}


We studied how the spin splittings $\Delta E$ for $N=1,2,3$ increase
with magnetic field from $B=0\; {\rm T}$ up to 9~T
(Fig.~\ref{FIG:TransconductanceAndEnergySplitting}b,d,f). We first
concentrate on data for $N=1$. At zero field $\Delta E$ shows the
splitting associated with the 0.7 anomaly, that we label $\Delta
E_{0.7}$. It is observed in all our QPCs with a typical value of
0.5~meV. At high fields $\Delta E$ has a linear slope similar to the
Zeeman effect. However, linear extrapolation of this slope down to
$B = 0$ shows that there is a large positive offset (unlike the
usual Zeeman effect). We characterize the slope with an effective
g-factor $|g^*| =\frac{1}{\mu_{B}} \frac{{\rm d}\Delta E}{{\rm d}B}$
(note that one should be careful to interpret $|g^*|$ as an absolute
indication for the g-factor of electrons in a QPC, since different
methods for extracting a g-factor can give different results
\cite{Cronenwett2002,Cronenwett2001}). The high-field offset from a
linear Zeeman effect is characterized with a parameter $\Delta
E_{hfo}$. Qualitatively, this type of data for $\Delta E$ looks
similar for all 12 QPCs
(Figs.~\ref{FIG:TransconductanceAndEnergySplitting}b,d,f), and we
use a suitable fitting procedure to characterize the traces for
$N=1$ with the parameters $\Delta E_{0.7}$, $\Delta E_{hfo}$ and
$|g^*|$. Notably, two-parameter fitting using spin-$\frac{1}{2}$
energy eigenvalues with $\Delta E = \sqrt{ (\Delta E_{0.7})^2 +(
|g^*| \mu_B B )^2}$ does not yield good fits. For the traces as in
Fig.~\ref{FIG:TransconductanceAndEnergySplitting}b,d,f for $N=2,3$,
we cannot resolve a spin splitting at low fields, only the
parameters $\Delta E_{hfo}$ and $|g^*|$ can be derived. Further
analysis of these results for $\Delta E_{0.7}$, $\Delta E_{hfo}$ and
$|g^*|$ is presented in Section~\ref{Sect:ManyBodyEffects} on
many-body effects.

\subsection{Electrostatics and subband splitting}

Figures~\ref{FIG:PinchOffAndSubbandSpacing}a,b present how the
pinch-off voltage $V_{po}$ (the value of $V_g$ where the $G$ starts
to increase from zero) depends on QPC geometry.
Figs.~\ref{FIG:PinchOffAndSubbandSpacing}c,d presents this for the
subband spacing $\hbar \omega_{12}$, which provides a parameter for
the strength of the transverse confinement in the QPC and is
possibly of importance for several of the many-body effects in QPCs.
The observed dependence of $V_{po}$ on the QPC geometry (a more
negative $V_{po}$ for shorter and wider QPCs) agrees with the
expected trend. This provides the first of several indications that
part of the physics of our set of QPCs depends in a regular manner
on $L$ and $W$.

Furthermore, the variation of $\hbar \omega_{12}$ is in good
agreement with an electrostatic analysis \cite{Davies1995} of the
degree of transverse confinement in the saddle-point-like potential
of the QPC (presented in
Figs.~\ref{FIG:PinchOffAndSubbandSpacing}e,f). In summary, short and
narrow QPCs yield the strongest transverse confinement
(Figs.~\ref{FIG:PinchOffAndSubbandSpacing}d-f). This is valid down
to the point where the QPC width $W \lesssim 3d$ (where $d$ the
depth of the 2DEG below the wafer surface), which results in the
maximum for $W = 350 \; {\rm nm}$ in
Figs.~\ref{FIG:PinchOffAndSubbandSpacing}d,f.

For this analysis, we calculate the confining electrostatic
potential due to the depletion gates in the plane of the 2DEG. An
important ingredient of the calculation is the threshold voltage
$V_t$, which is the (negative) voltage that must be applied to a
gate to reduce the electron density underneath it to zero,
\begin{equation}
V_t = -\frac{e n_{2D} d}{\epsilon_r \epsilon_0},
\end{equation}
where $n_{2D}$ is the 2DEG electron density (without gates) and $d$
the depth of the 2DEG below the surface. We use the dielectric
constant for GaAs $\epsilon = \epsilon_r \epsilon_0 = 12.9$.

The subband spacing can be calculated from the transverse curvature
of the saddle-point-like potential. However, this curvature changes
with the applied gate voltage. Therefore, we calculate the curvature
for all QPCs when the QPC is just pinched-off ($G=0$), when the
potential in the middle of the transport channel is equal to $V_t$.
For these calculations we used the dimensions $L$, $W$ and $d$ of
the measured devices. Qualitatively the trends in $\hbar
\omega_{12}$ as a function of $L$ and $W$ are reproduced, but the
calculated values for $\hbar \omega_{12}$ are significantly larger
than the experimentally obtained values. From earlier work
\cite{Davies1995} it is known that it is hard to get quantitative
agreement from this type of calculations. Furthermore, the maximum
in $\hbar \omega_{12}$ versus $W$ is found for a smaller value for
$W$ than we have observed experimentally, approximately when $W \sim
2d$.

The origin of differences between our simple calculations and
experimental results is well understood. The treatment of the
exposed surfaces between the depletion gates is an important aspect
of the calculations. A different choice for the boundary condition
of the exposed surface may result in a noticeable difference in the
confining potential \cite{Davies1995,Chen1993}. We used a so-called
pinned-surface approach (because it is a simple analytical approach)
where the Fermi level at the surface becomes pinned to the Fermi
level in the 2DEG. However, a pinned surface requires charge to move
from the 2DEG to the surface when the gate voltage is changed, in
order to keep the surface potential constant. This process is
strongly suppressed at low temperatures. Alternatively, the surface
can be treated as a dielectric boundary, with a fixed charge density
(frozen surface approach). Davies \textit{et al.} \cite{Davies1995}
have compared the results for pinned and frozen surfaces and found
that the maximum in $\hbar \omega_{12}$ shifts from $W \sim 2d$ to
$W \sim 3d$ when a frozen surface is assumed instead of a pinned
surface. This corresponds very well to the experimentally observed
value of 350 nm. Furthermore, the model used here is based on the
calculation of the electrostatic potential due to the gates alone.
Other effects, as the contribution to the potential from donor ions
and other electrons in the 2DEG are ignored. Self-consistent
calculations \cite{Laux1988} have shown that the values of $\hbar
\omega_{12}$ decrease rapidly when electrons enter the conduction
channel.

\section{Many-body effects} \label{Sect:ManyBodyEffects}

\subsection{Enhancement of the effective g-factor}

\begin{figure}
  \includegraphics[width=1\columnwidth]{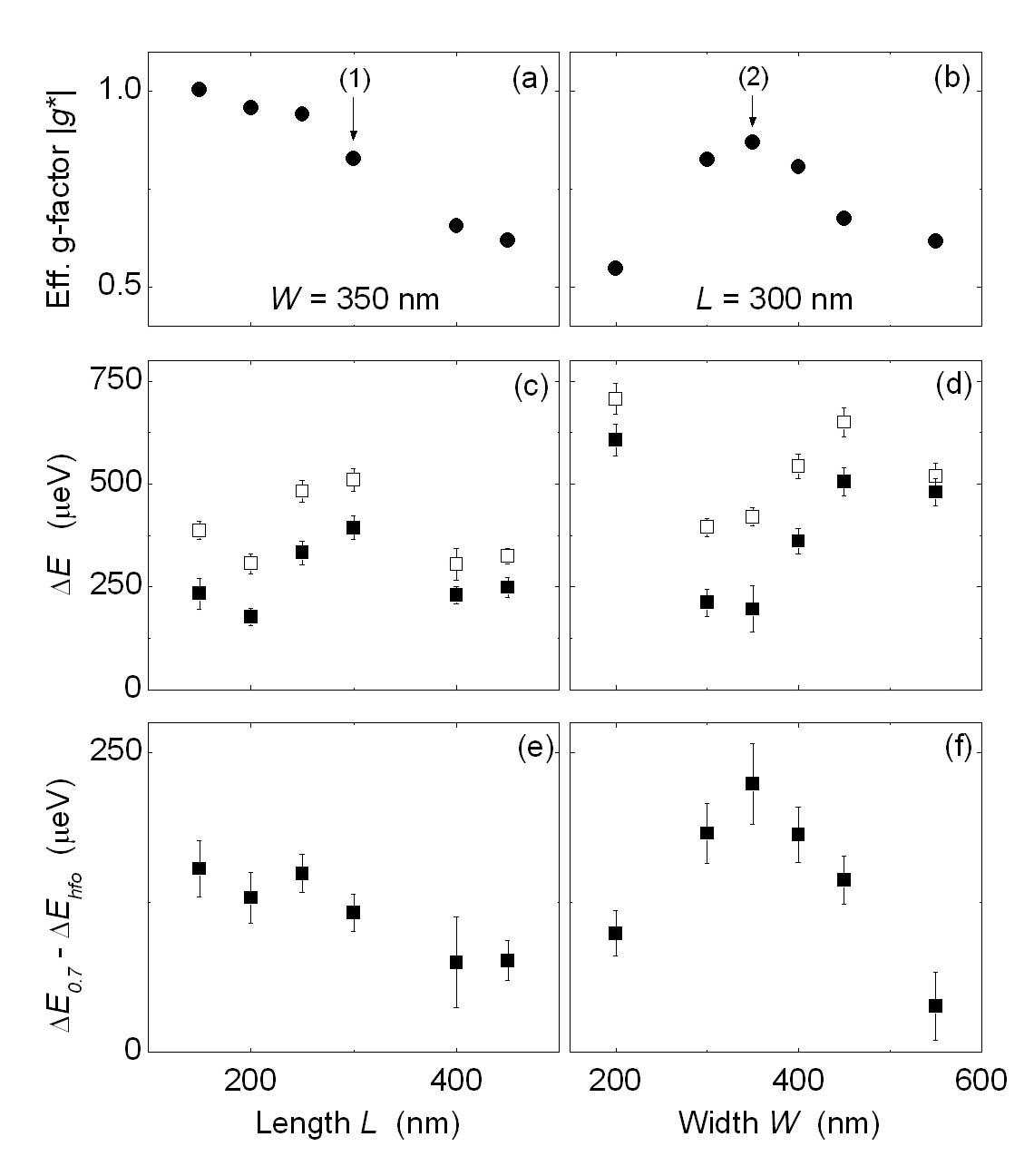}\\
  \caption{(a),(b)
Effective g-factor $|g^*|$ as a function of QPC
  length $L$ (with fixed width $W$ = 350 nm), and as a function QPC width $W$ (with
  fixed length $L$ = 300 nm). The effective g-factor $|g^*|$ is enhanced as compared to the bulk 2DEG value
  (up to a factor $\sim 3$) and shows a clear dependence on $L$ and $W$, that is correlated with the dependence of the subband spacing $\hbar
  \omega_{12}$ in Fig.~\ref{FIG:PinchOffAndSubbandSpacing}c,d.
  Data points labeled with (1) and (2) are from two different devices
with nominally identical values for $L$ and $W$ (see text for
details).  (c),(d) The 0.7 energy splitting $\Delta E_{0.7}$ and
high-field offset $\Delta E_{hfo}$ for the $N=1$ subband as a
function of $L$ and $W$. The values of
  $\Delta E_{0.7}$ and $\Delta E_{hfo}$ both vary with $L$ and $W$ in a irregular manner,
  but there is a strong correlation between $\Delta E_{0.7}$ and $\Delta
  E_{hfo}$. (e),(f) The difference between $\Delta E_{0.7}$ and $\Delta
  E_{hfo}$ as a function of $L$ and $W$. This data again shows a correlation with the dependence
of subband spacing $\hbar \omega_{12}$ on $L$ and $W$. All data
points are for the $N=1$ subband from results measured
  at 200~mK.
(Fig.~\ref{FIG:PinchOffAndSubbandSpacing}c,d).}
 \label{FIG:GfactorAndEnergySplittingVSShape}
\end{figure}

Figures~\ref{FIG:GfactorAndEnergySplittingVSShape}a,b present how
the effective g-factor $|g^*|$ for $N=1$ varies with $L$ and $W$ of
the QPCs. It is strongly enhanced up to a factor $\sim 3$ with
respect to the g-factor for bulk 2DEG material \cite{Hannak1995}
(the temperature dependence of this $|g^*|$ data is shortly
discussed in section~\ref{SubSect:AnomalyExhange}). This has been
observed before \cite{Thomas1998} and is attributed to many-body
effects. Notably, the values of $|g^*|$ and $\hbar \omega_{12}$ in
Figs.~\ref{FIG:PinchOffAndSubbandSpacing}c,d are clearly correlated.

The enhancement of the effective g-factor has been explained in
terms of exchange interactions (see \cite{Pallecchi2002} and
references therein). Calculations of the exchange potential in a
square (quantum well) confining potential have shown that the
effective g-factor decreases when the 1D confining potential weakens
and the 2D limit is approached \cite{Majumdar1998}. For a harmonic
confining potential, the results of this study predict that $|g^*|$
scales indeed with $\hbar \omega_{12}$. We observed this here for
the lowest subband ($N=1$) in dependence on QPC geometry. Earlier
work \cite{Thomas1998} observed the same trend in a single QPC,
using that the transverse confinement decreases with increasing
subband index $N$.

\subsection{The 0.7 anomaly and exchange} \label{SubSect:AnomalyExhange}

Figures~\ref{FIG:TransconductanceAndEnergySplitting}b,d,f show that
for all QPCs $\Delta E$ appears in high fields as the sum of the
Zeeman effect and the constant contribution $\Delta E_{hfo}$. This
suggest that the splittings in high field have, in particular for
$N=1$, a significant contribution from a field-independent exchange
effect that results from each subband being in a ferromagnetic
spin-polarized state. In high fields such an interpretation is less
ambiguous than for zero field (where the possibility of a
ferromagnetic ground state for spin-polarized subbands is the topic
of debate \cite{Klironomos2006,Jaksch2006}) since the Zeeman effect
suppresses spin fluctuations. Thus, measuring $\Delta E_{hfo}$ can
be used to determine this exchange splitting.

We now further analyze how this parameter $\Delta E_{hfo}$ and
$\Delta E_{0.7}$ depend on $L$ and $W$. The open squares in
Figs.~\ref{FIG:GfactorAndEnergySplittingVSShape}c,d present this for
$\Delta E_{0.7}$. Overall, the dependence here is not very regular,
possibly indicating that the exact appearance of the otherwise
robust 0.7 anomaly is sensitive to small irregularities in the
potential that defines the QPC (only the data in
Fig.~\ref{FIG:GfactorAndEnergySplittingVSShape}d suggests an
anti-correlation with $\hbar \omega_{12}$). The black squares
present how $\Delta E_{hfo}$ for $N=1$ varies with $L$ and $W$. Also
here the dependence is irregular. Remarkably, however, the irregular
variations of $\Delta E_{0.7}$ and $\Delta E_{hfo}$ are clearly
correlated throughout our set of 12 QPCs. This means that $\Delta
E_{0.7}$, which is derived from data in zero field, is correlated
with $\Delta E_{hfo}$, which is derived from data taken at fields in
excess of 5~T. Further evidence for the significance of this
correlation comes from data from the $N=2$ and $N=3$ subband (see
Figs.~\ref{FIG:TransconductanceAndEnergySplitting}b,d,f). We
analyzed the data for $N=2,3$ in the very same way as for $N=1$, and
the most important observation is that the $\Delta E_{hfo}$
parameter for $N=2,3$ is much smaller than for $N=1$, and often
close to zero. A high $\Delta E_{hfo}$ value is only observed for
$N=1$, just as the 0.7 anomaly itself. Notably, for $N=1$, $\Delta
E_{0.7}$ and $\Delta E_{hfo}$ also have a similar order of
magnitude. This analysis points to the conclusion that the
spontaneous energy splitting of the 0.7 anomaly is dominated by the
same effect that causes the high-field offset $\Delta E_{hfo}$. As
we discussed, this is probably an exchange contribution
\cite{Koop2007}. The error bar that we attribute to these values
includes an error from the transconductance peak-fitting, one from
the conversion of gate voltage to energy scale, and an error due to
scatter in the $\Delta E$ datapoints as a function of $B$.

Fig.~\ref{FIG:GfactorAndEnergySplittingVSShape}e,f presents data for
the difference between $\Delta E_{0.7}$ and $\Delta E_{hfo}$. Here,
$\Delta E_{0.7} - \Delta E_{hfo}$ shows again a correlation with
$\hbar \omega_{12}$. This indicates that the splitting of the 0.7
anomaly has (in addition to the exchange contribution that is also
present in high fields) a contribution that scales with $\hbar
\omega_{12}$. At this stage we cannot relate this new observation to
earlier experimental or theoretical work. Note that for the error
bars in Figs. \ref{FIG:GfactorAndEnergySplittingVSShape}e,f we first
subtracted the values of peak positions in terms of gate voltage,
such that the error from gate voltage to energy scale conversion is
accounted for only once.

\begin{figure*}
  \includegraphics[width=1.9\columnwidth]{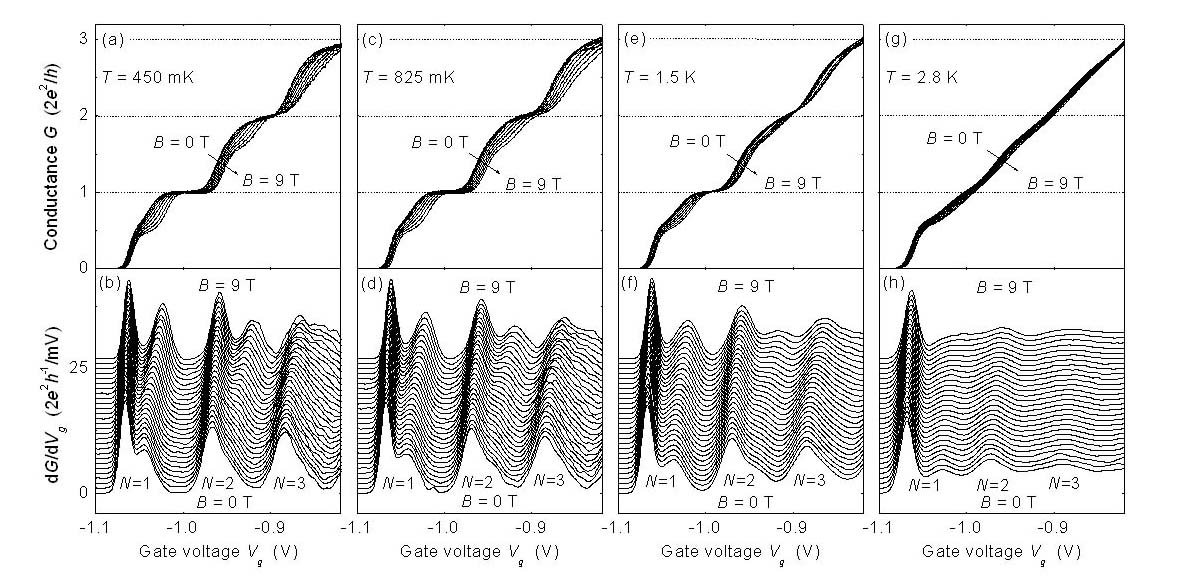}\\
  \caption{(a)
Differential conductance $G$ as a function of gate voltage $V_g$ at
450 mK, for a QPC with $L$ = 300 nm and $W$ = 400 nm. The in-plane
magnetic field is increased from $B = 0 \; {\rm T}$ to $B = 9 \;
{\rm T}$. (b) Transconductance ${\rm d}G / {\rm d}V_g$
  traces (offset vertically for clarity) obtained by differentiating the data in (a). The conductance $G$ and transconductance ${\rm d}G / {\rm d}V_g$ as in (a),(b) are shown for $T$ = 825
  mK in (c),(d), for $T$ = 1.5 K in (e),(f) and for $T$ = 2.8 K in (g),(h).}
  \label{FIG:TransconductanceTempDependence}
\end{figure*}

We will now discuss the effect of increasing the temperature on the
many-body phenomena in our QPCs.
Figs.~\ref{FIG:TransconductanceTempDependence}a-d show the
conductance $G$ at $V_{sd} \sim 0$ as a function of magnetic field
for temperatures $T$~=~450~mK, 825~mK, 1.5~K and 2.8~K (see also
Figs.~\ref{FIG:DeviceAndConductance}a and
\ref{FIG:TransconductanceAndEnergySplitting}a for the 200~mK data).
As the temperature is increased the spin-degenerate plateaus and the
spin-resolved plateaus both become less pronounced due to thermal
smearing. In high magnetic fields the spin-resolved plateaus
increase slightly in conductance with increasing temperature. At
even higher temperatures the plateau at $0.7(2e^2/h)$ is the last
remaining feature in the differential conductance. Notably, here the
0.7 anomaly appears to be present over the whole range of magnetic
fields. The corresponding transconductance traces ${\rm d}G / {\rm
d}V_g$ are plotted in
Figs.~\ref{FIG:TransconductanceTempDependence}e-h. As a result of
the thermal smearing of the conductance plateaus, the peaks in ${\rm
d}G / {\rm d}V_g$ become broader and decrease in height. The
zero-field splitting in the transconductance peak for $N = 1$ has
been identified as the 0.7 anomaly. When the temperature is
increased, the 0.7 anomaly becomes more pronounced as was shown in
the temperature dependence of the differential conductance $G$
presented in Fig. \ref{FIG:DeviceAndConductance}d. Consequently the
zero-field splitting in
Figs.~\ref{FIG:TransconductanceTempDependence}e-h also increases.
For $T$ = 825 mK and 1.5 K (Figs.
\ref{FIG:TransconductanceTempDependence}f,g) even a small zero-field
splitting of the $N = 2$ transconductance peak can be observed,
suggesting the appearance of a $1.7(2e^2/h)$ plateau
\cite{Thomas1998}.

\begin{figure}[h!]
  \includegraphics[width=0.8\columnwidth]{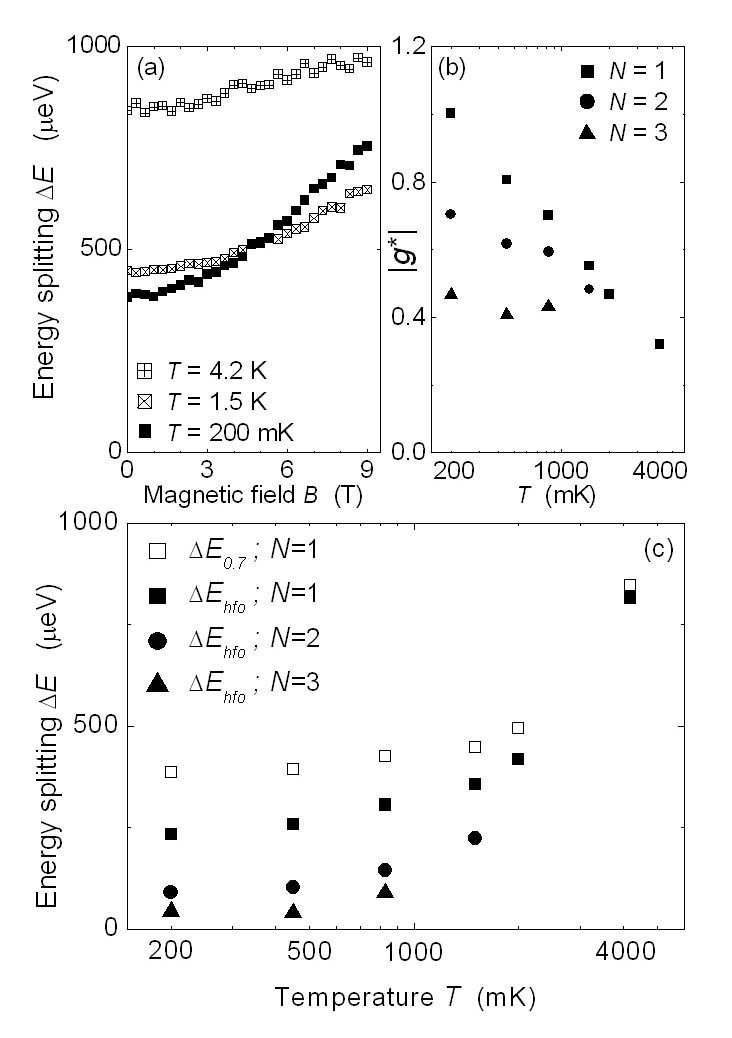}\\
  \caption{(a) Energy splitting for $N=1$ as a function of magnetic field for
  different temperatures $T$, for a QPC with $L$ = 150 nm and $W$ = 350
  nm. (b) Effective g-factor $|g^*|$ as a function of temperature for the same QPC. The
  g-factor enhancement is strongest for the $N=1$ subband at the
  lowest temperature. For the $N=2$ and $N=3$ subband the g-factor
  is also enhanced at low temperatures. As the temperature is
  increased the g-factor enhancement is weaker for all subbands. (c) The 0.7 energy
  splitting $\Delta E_{0.7}$ and high-field offset $\Delta E_{hfo}$ as a function of temperature. The value for $\Delta
  E_{hfo}$ is highest for the $N=1$ subband and decreases to zero with increasing
  subband number. As the
  temperature is increased, the $\Delta E_{0.7}$ value as well as the $\Delta E_{hfo}$ values for $N=1,2,3$ strongly increase.
  The correlation between $\Delta E_{0.7}$ and $\Delta E_{hfo}$ remains present upon increasing the temperature.}
  \label{FIG:EnergySplittingVSTemperature}
\end{figure}

Using the temperature dependence of $\Delta E$ data
(Fig.~\ref{FIG:EnergySplittingVSTemperature}a), we find that the
correlation between $\Delta E_{0.7}$ and $\Delta E_{hfo}$ remains
intact at higher temperatures
(Fig.~\ref{FIG:EnergySplittingVSTemperature}c).
Figure~\ref{FIG:EnergySplittingVSTemperature}b shows that $|g^*|$
has a very different temperature dependence. This indicates that the
g-factor enhancement and the 0.7 anomaly arise from different
many-body effects.

\subsection{Kondo signatures}

The appearance of the 0.7 anomaly has been related to a peak in the
differential conductance as a function of source-drain voltage
around zero bias, for $G$ values around $e^2/h$. Earlier work
\cite{Cronenwett2002} showed that this zero-bias anomaly (ZBA), and
its temperature and magnetic field dependence, have a very striking
similarity with electron transport through a Kondo impurity that can
studied with quantum dots \cite{GoldhaberGordon1998,Cronenwett1998}.
For quantum dots, the Kondo effect is a many-body interaction of the
localized electron(s) inside the dot with the delocalized electrons
in the leads connected to the dot
\cite{GoldhaberGordon1998,Kouwenhoven2001,Cronenwett1998,Cronenwett2002}.
Together these electrons form a spin-singlet state, effectively
screening the local spin on the dot. In contrast to a quantum dot,
where there is a clear localized state, a QPC is an open system
where the formation of a bound state is much less obvious. A recent
theoretical result \cite{Rejec2006} has shown that a self-consistent
many-body state can indeed form inside a QPC, and that this can
result in Kondo-like physics.

\begin{figure}[h!]
  \includegraphics[width=1\columnwidth]{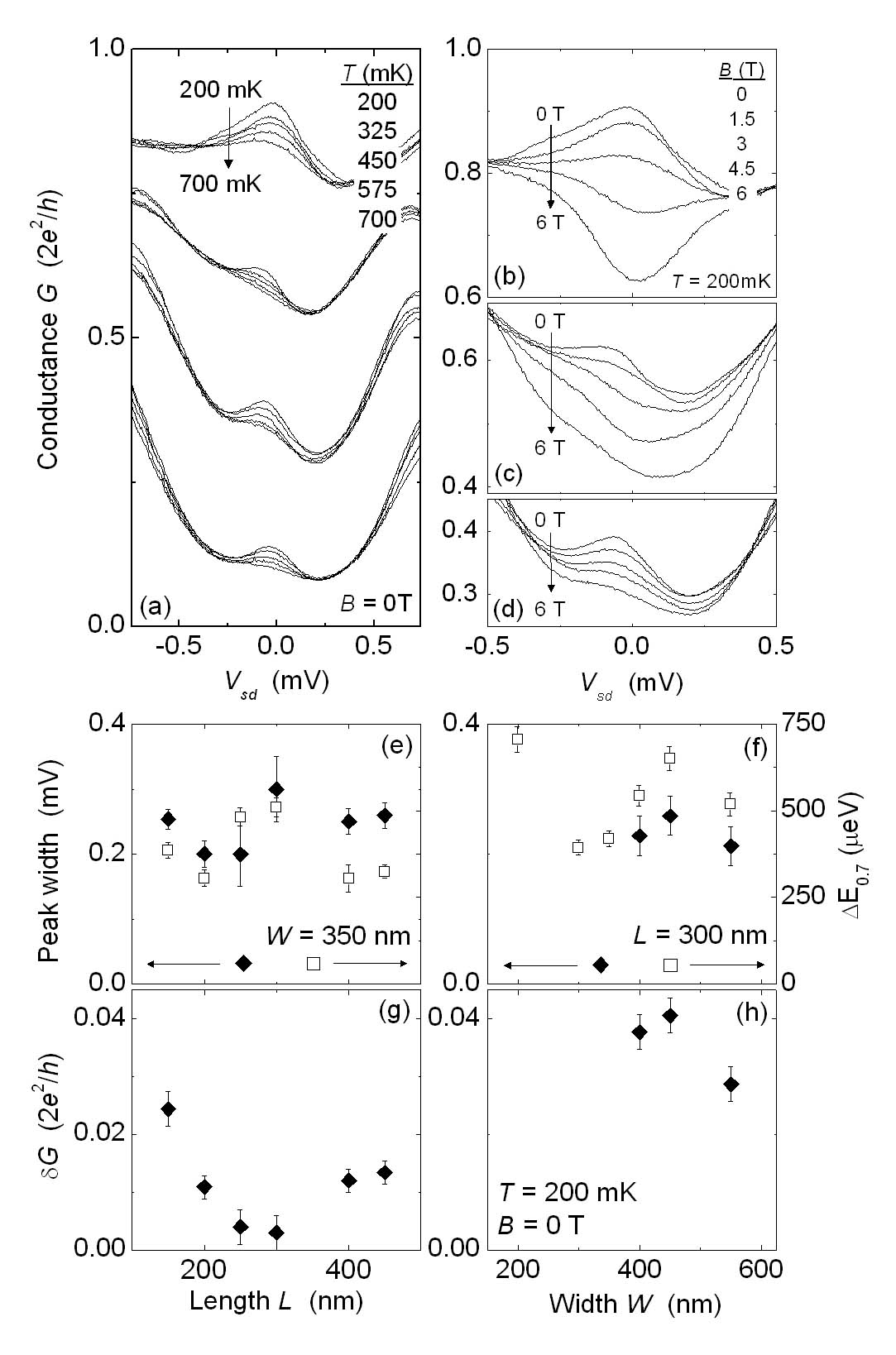}\\
  \caption{(a) Temperature dependence of the zero-bias anomaly (ZBA) for 4 fixed values of the gate
voltage. The peak becomes more pronounced as temperature is lowered
from 700 to 200 mK. (b-d) Magnetic field dependence of the ZBA for
$G \sim 0.8$ (b), $G \sim 0.5$ (c) and $G \sim 0.3(2e^2/h)$ (d). The
ZBA should split by upon application of an in-plane magnetic field
\cite{Cronenwett1998,Cronenwett2002}. The peak in (b) does not split
but collapses with $B$ because $T_K~<~2g^*\mu_BB$ in this regime
  \cite{Cronenwett2002}. The splitting in (c) and (d) is not very
  prominent, possibly due to our electron temperature of $\sim$ 200 mK. (e),(f)
  The width of the ZBA at $G \sim 0.3(2e^2/h)$ (left axis) as a function
  of QPC length $L$ and width $W$. The peaks are fitted in zero
  magnetic field at $T$ = 200 mK. The right axis shows $\Delta E_{0.7}$
  (data from Figs. \ref{FIG:GfactorAndEnergySplittingVSShape}g,h) (g),(h) Height of the ZBA also at $G
  \sim 0.3(2e^2/h)$ versus $L$ and $W$ for the same conditions as in (e),(f).}
  \label{FIG:KondoSignatures}
\end{figure}

In this section we present the measurements of this ZBA in our set
of QPCs. Most of our QPCs showed a clear ZBA in nonlinear
conductance measurements. The temperature and magnetic field
dependence of this data (Figs.~\ref{FIG:KondoSignatures}a-d) is
consistent with the earlier reports \cite{Cronenwett2002} that
relate the 0.7 anomaly to transport through a Kondo impurity.

The relevant energy scale for Kondo physics is the Kondo temperature
$T_K$. Below this temperature the magnetic impurity giving rise to
Kondo physics is completely screened by the formation of a
spin-singlet state and at zero-bias the differential conductance $G
\cong 2e^{2}/h$. The Kondo temperature determines the width of the
zero-bias peak. We observe that the peak width and height $\delta G$
of the ZBA are not constant over the whole range 0 $<$ G $<$
$2e^2/h$ (see also Fig. 6-12 in reference \cite{Cronenwett2001}). We
choose to fit these parameters at $G \sim$ 0.3, where the peak
height has a relative maximum. The peak width and height are
determined by fitting the nonlinear conductance traces with a
Gaussian shaped peak added to a parabola.

Figures~\ref{FIG:KondoSignatures}e,f show the peak width as a
function of $L$ and $W$ (during our measurement run, one gate of
Device 2 broke during an electronic malfunction, and we can only
present data from 3 QPCs in the set with different values of $W$).
The width of the ZBA does not show a clear dependence on $L$ and
$W$, and has a value of about 2 mV for all QPCs. For completeness,
we also report the peak height $\delta G$ in
Figs.~\ref{FIG:KondoSignatures}g,h as a function of $L$ and $W$. We
observe that for a single QPC $\delta G$ varies with $V_g$, but the
the values in Figs.~\ref{FIG:KondoSignatures}g,h do give for each
QPC a good representation of the typical value of $\delta G$
throughout the $V_g$ interval where the ZBA is observed. As a
function of $L$ and $W$, we observe here a stronger scatter in the
values than for the peak width, but also here there is no clear
relation with the QPC geometry. To conclude this section, we
consider a correlation between the signatures of the Kondo effect
and the values of $\Delta E_{0.7}$
(Figs.~\ref{FIG:KondoSignatures}e,f). The irregular variation of the
ZBA peak width with $L$ and $W$ suggests indeed a correlation with
$\Delta E_{0.7}$, but here the evidence is very weak given the size
of the error bar that we attribute to these values.

\section{Discussion and conclusions} \label{Sect:Conclusions}

We have studied many-body interaction effects in quantum point
contacts. Our main point of interest was the dependence of these
many-body electron interactions on the geometry of the QPC. We found
a clear relation between the subband spacing and the enhancement of
the effective electron g-factor. These parameters depend on geometry
in a regular manner that we can understand from electrostatic
modeling of the QPC potential. The many-body electron physics that
causes the spontaneous energy splitting of the 0.7 anomaly does not
show a clear dependence on QPC geometry, but we do find a clear
correlation with a field-independent exchange effect that
contributes to spin splittings in high magnetic fields. This
suggests that the splitting of the 0.7 anomaly is dominated by this
exchange contribution. We also measured a zero-bias anomaly  in the
nonlinear conductance of our QPCs, that has been interpreted as a
signature of the Kondo effect. Here, there is also no clear
dependence on QPC geometry, but our data suggests that it is
worthwhile to further study its correlation with the splitting of
the 0.7 anomaly. These results are important for theory work that
aims at developing a consistent picture of many-body effects in
QPCs, and its consequences for transport of spin-polarized electrons
and spin coherence in nanodevices. Our analysis of experimental data
is very phenomenological, presenting parameters and correlations for
which it is difficult to draw conclusions about the underlying
physics. At the same time, part of state-of-the-art theory work now
relies on numerical simulations of realistic QPC geometries (using
spin-density-functional theory \cite{Rejec2006,Jaksch2006} or other
numerical approaches \cite{Klironomos2006}) from which it is hard to
derive analytical expression for the underlying physics. However,
the validity of this numerical modeling can be easily tested for its
consistency with the parameters and correlations that we reported
here.


\vskip 5mm

\noindent {\bf Acknowledgements}

We thank B.~H.~J.~Wolfs, R.~N.~Schouten, S.~F.~Fischer,
C.~W.~J.~Beenakker and L.~S.~Levitov for help and useful
discussions, and the Dutch Foundation for Fundamental Research on
Matter (FOM), the Netherlands Organization for Scientific Research
(NWO), and the German BMBF (in the framework of the
nanoQUIT-program) for financial support.


\end{document}